\documentclass[aps,superscriptaddress,twocolumn,pra,longbibliography]{revtex4-1}
\usepackage[usenames,dvipsnames]{xcolor}
\usepackage{graphicx}
\usepackage{color}
\usepackage[usenames,dvipsnames]{xcolor}
\usepackage{amsmath,bbm,amssymb, amsthm,natbib}
\usepackage{dsfont} 
\usepackage{stmaryrd}
\usepackage{numprint}
\begin{document}
	\title{Superconducting quantum refrigerator: Breaking and rejoining Cooper pairs with magnetic field cycles} \author{Sreenath K. Manikandan}
	\email{skizhakk@ur.rochester.edu}
	\affiliation{Department of Physics and Astronomy, University of Rochester, Rochester, NY 14627, USA}
	\affiliation{Center for Coherence and Quantum Optics, University of Rochester, Rochester, NY 14627, USA}   
	\author{Francesco Giazotto}
		\email{francesco.giazotto@sns.it}
	\affiliation{NEST, Istituto Nanoscienze-CNR and Scuola Normale Superiore, Piazza San Silvestro 12, 56127 Pisa, Italy}

	\author{Andrew N. Jordan}
	\email{jordan@pas.rochester.edu}
	\affiliation{Department of Physics and Astronomy, University of Rochester, Rochester, NY 14627, USA}
	\affiliation{Center for Coherence and Quantum Optics, University of Rochester, Rochester, NY 14627, USA}    
	\affiliation{Institute for Quantum Studies, Chapman University, Orange, CA, 92866, USA}
	\date{\today}
	
	\begin{abstract}
We propose a solid state refrigeration technique based on repeated adiabatic magnetization/demagnetization cycles of a superconductor which acts as the working substance. The gradual cooling down of a substrate (normal metal) in contact with the working substance is demonstrated for different initial temperatures of the substrate. Excess heat is given to a hot large-gap superconductor.  The on-chip refrigerator works in a cyclic manner because of an effective thermal switching mechanism: Heat transport between N/N versus N/S junctions is asymmetric because of the appearance of the energy gap.  This switch permits selective cooling of the metal. We find that this refrigeration technique can cool down a 0.3cm$^{3}$ block of Cu by almost two orders of magnitude starting from 200mK, and down to about 1mK starting from the base temperature of a dilution fridge (10mK). The corresponding cooling power at 200mK and 10mK for a 1cm$\times$1cm interface are 25 nW and 0.06 nW respectively, which scales with the area of the interface.
	\end{abstract}
		\maketitle
The goal of building solid state refrigerators and heat engines working on quantum principles is an outstanding need for next generation quantum technologies~\cite{sothmann2014thermoelectric}. It is known since the earlier days of superconductivity that the process of magnetizing a superconducting material quasistatically and adiabatically can reduce the temperature of the material substantially as it transitions to the normal state~\cite{keesom1934further,mendelssohn1934magneto,yaqub1960cooling}.  This is because a material in its superconducting state has more order, and therefore, entropy equal to that of a normal metal at a lower temperature. Hence when driven to the normal state adiabatically by an applied magnetic field, the achieved final state is much colder than the initial superconducting state as depicted in the $T-\mathcal{S}$ diagram in Fig.~1(b). There were attempts in the past to try and implement adiabatic magnetization of a superconductor as an effective cooling technique, notably the early proposals by Mendelssohn and Moore~\cite{mendelssohn1934magneto}, and by Keesom and Kok~\cite{keesom1934further}. Recently Dolcini  and Giazotto had studied the adiabatic magnetization of a superconductor by including dynamical dissipative effects such as eddy current losses, and suggested that this mechanism can still be used to achieve significant cooling for micro-refrigeration purposes~\cite{dolcini2009adiabatic}.  
\begin{figure}
\includegraphics[width=\linewidth]{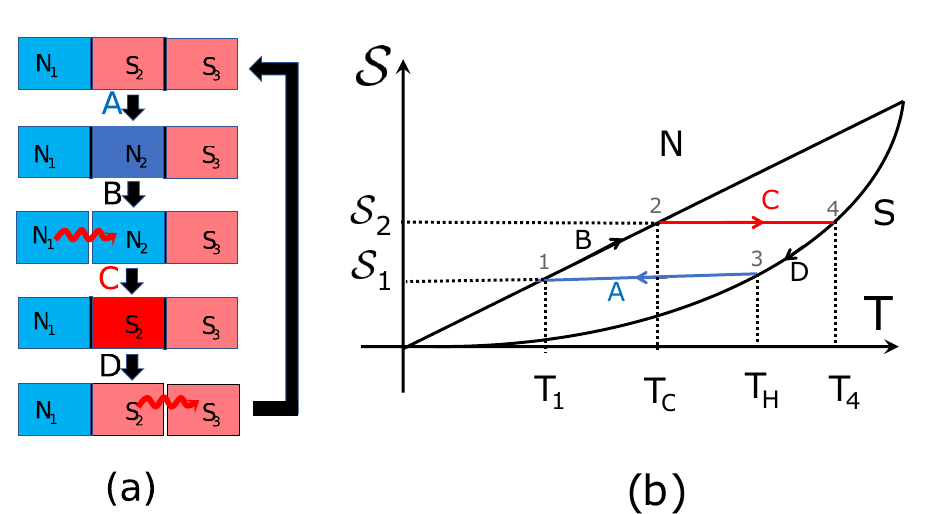}\caption{(a) Steps of the refrigeration cycle.  In step A, the central region (working substance) is thermally isolated from its neighbors, and undergoes adiabatic magnetization from a superconductor to a normal metal, $S_2 \rightarrow N_2$, and cools to a much colder temperature, $T_1$.  In step B, thermal contact with the normal metal $N_1$ is made, resulting in heat transfer from $N_1$ to $N_2$, eventually coming to equilibrium at temperature $T_C$.  In step C, the working substance is thermally isolated again (black walls), and adiabatically demagnetized from $N_2 \rightarrow S_2$, heating up the system to its hottest temperature $T_4$.  In step D, thermal contact with superconductor $S_3$ is made, allowing heat to escape from $S_2 \rightarrow S_3$, reducing the temperature to temperature $T_H$. The cycle closes by closing off thermal contact to $S_3$ with the black wall, and returning to step A. (b) Entropy of the normal metal and a superconductor, showing different stages of the refrigeration cycle.\label{fig1}}
\end{figure}
\begin{figure*}[htb!]
\includegraphics[width=\linewidth]{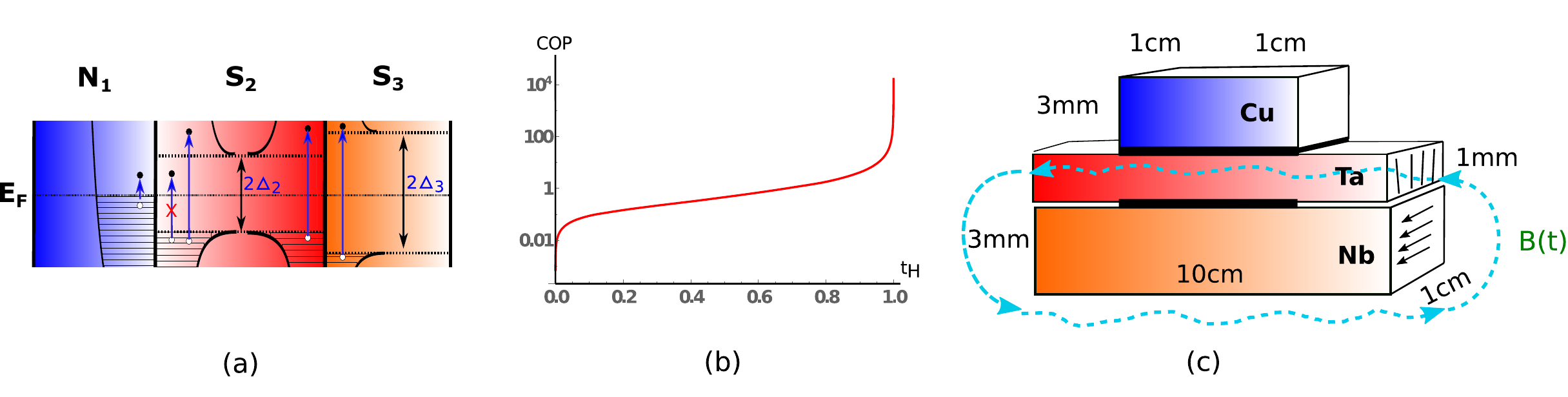}
\caption{ (a) The energy diagram of the junctions when the working substance is in the hot superconducting state. Thermal excitations which create quasi-particles carrying the heat flux are indicated by blue vertical arrows. The appearance of an energy gap in the hot superconducting phase of the working substance exponentially suppresses the reverse heat flux between the working substance and the cold reservoir, since quasi-particle excitations are forbidden below the gap. Further, coupling to the hot reservoir superconductor with a larger superconducting gap facilitates selective removal of high energy quasi-particles from the working substance. The gap energies, and the Fermi energy $\text{E}_{F}$ are marked in the diagram. (b) Coefficient of Performance (COP) of the proposed refrigerator, evaluated as a function of  $t_H=\frac{T_{H}}{T_{*}}$. We have set $T_{i}=T_{H}$. (c) Proposed architecture of the device. We consider Copper (Cu, $\gamma = 0.69$ mJ/(mol K$^{2}$), Debye temperature $\theta_{D}=$ 347 K) as the substrate, Tantalum (Ta,  $\gamma = 5.87$ mJ/(mol K$^{2}$), $\theta_{D}=$ 246 K) as the working substance and Niobium (Nb,  $\gamma = 7.80$ mJ/(mol K$^{2}$), $\theta_{D}=$ 276 K) as the hot reservoir~\cite{tari2003specific}. A super-current can be applied in Nb, which generates the magnetic field that drives the phase transition in the adjacent working substance.\label{fridge}}
\end{figure*}

Here we propose a cyclic superconducting refrigerator based on adiabatic magnetization of a superconductor, with a working mechanism similar to that of a domestic refrigerator. A conventional refrigerator operates by cyclically moving a working fluid between hot and cold reservoirs. Work is done by compressing a fluid, and letting it freely expand to a gas in a phase transition where it cools down and absorbs heat from the cold reservoir. The now hot gas is then re-compressed, liquifying it, and then dumps the excess heat to a hot reservoir, which is usually the environment that allows the fluid to thermalize and reset to its initial temperature. The cycle repeats many times such that a stable low final temperature is achieved in the cold reservoir.
In comparison, the working fluid in our example is the electron gas in the working superconductor. The cold reservoir is a normal metal, and the hot reservoir is another superconductor having a larger gap.  The superconducting state of electrons in the working substance is analogous to a compressed fluid. With an applied magnetic field, the electron fluid expands in a phase transition into the unpaired normal state at a lower temperature. Heat is then absorbed from the cold reservoir, and the electron fluid is re-compressed by reducing the applied magnetic field. The working substance, which is now hotter than the hot reservoir, has reduced electronic entropy in the paired state. The entropy of phonons has increased in the working substance in proportion, effectively holding the excess heat. Note that here the phonon entropy changes complementary to the electron entropy (in both steps, A and C), such that the sum of the entropy of electrons and phonons remains constant, and the process is adiabatic. Electron-phonon interactions in the working substance and a tunneling contact with the hot reservoir selectively removes hot electrons from the working substance, and facilitate reaching thermal equilibrium in the hot junction. This cycle repeats, establishing a low temperature steady state in the cold reservoir. 

\textit{The cyclic superconducting refrigerator} -- Adiabatic magnetization of a superconductor preserves the total entropy of the material such that the entropies of the two phases are equal,
    $\mathcal{S}^{N}(T_{f}, H=H_{c}) = \mathcal{S}^{S}(T_{i}, H=0)$, where $H$ is the applied magnetic field.
This results in cooling of the material to a final temperature $T_{f}$ that is approximately equal to $T_{i}^{3}/T_{*}^{2}$. Here $T_{*}^{2} = \frac{\gamma_{2}}{\alpha_{2}}$, is a characteristic temperature of the working substance~\cite{yaqub1960cooling,dolcini2009adiabatic}. We consider Tantalum as the working substance for which $T_{*}=11.6$K.

 For the cyclic superconducting refrigerator presented here, the crucial point is that the magnetic field inducing the phase transition, when applied quasi-statically, can be reversed quasi-statically to its initial value and therefore reversing the superconducting to normal phase transition of the working substance. This cycle can be performed repeatedly, where the working substance is driven between two different temperatures (hot and cold), envisaging a refrigeration cycle. The energy transfer is asymmetric.  That is, energy flow has a preferred direction that is different for the different phases, as a consequence of the energy-structure of the N/S materials~\cite{martinez2015rectification,giazotto2013thermal,martinez2013efficient}. The proposed refrigerator is sketched in Fig.~\ref{fig1} and Fig.~\ref{fridge} (a). 

We assume that the initial temperature of the working substance $T< 0.1~T_{c}$ of the working substance, such that its specific heat in the superconducting state can be approximated by~\cite{yaqub1960cooling},
 \begin{equation}
        C_{S} = 3\alpha T^3 + a \gamma T_{c}\exp (-b T_{c}/T),\end{equation}
where $b=1.44$, and $a=9.14$. Here $\alpha$ and $\gamma$ are parameters specifying its specific heat at the normal state, 
 $   C_{N} = 3\alpha T^3 + \gamma T.$
Here the common $T^3$ term is the phononic (Debye) contribution to the specific heat~\footnote{The phononic contribution to specific heat is related to the Debye temperature $\theta_{D}$ by the relation, $C_{ph}(T)=1944\frac{T^{3}}{\theta_{D}^{3}}$J/(mol K)~\cite{pobell1996matter}}. In the superconducting case, the exponential behavior of the electronic specific heat of the superconductor at low temperatures can be associated to the presence of a superconducting gap. The critical field as a function of temperature can be found from free energy differences~\cite{dolcini2009adiabatic}, which agrees reasonably well with the empirical formula, $H_{c}(T) = H_{0}(1-\frac{T^{2}}{T_{c}^{2}})$ for $T<T_{c}$, where $H_{0}$ is the zero-temperature critical field of the working substance. We consider $B_{0}=\mu_{0}H_{0} \simeq 0.08$T as the critical magnetic field at zero temperature for our working substance, Tantalum~\cite{milne1961superconducting}.

We can calculate some ideal thermodynamic properties of the cyclic refrigerator. The temperature of the  hot reservoir is $T_{H}$. The working substance is in thermal equilibrium with the hot reservoir initially and the following cycle occurs  (See Fig.~\ref{fig1} and Fig.~\ref{fridge}):
\begin{itemize}
    \item {Step A: A quasistatically applied magnetic field  drives the working substance to the normal state. The transformation is iso-entropic and the working substance cools down to $T_{1}= T_{H}^{3}/T_{*}^{2}$. Magnetic work $W_{3,1}$ is done.}
    \item{Step B: The working substance is put in contact with the cold reservoir where it absorbs heat $Q_{C}$. Since the electronic contribution to the entropy and specific heat dominates in the normal state, the transferred heat per unit volume can be approximated,}
  $  Q_C = \int T d\mathcal{S} = \frac{\gamma_{2}}{2} (T_C^2 - T_1^2).$
\end{itemize}
The temperature $T_{C}$ can be identified as the equilibrium final temperature between the cold reservoir and the working substance, approximated as $T_{C} \simeq \sqrt{\frac{\gamma_{2}T_{1}^{2}+\gamma_{1}T_{i}^2}{\gamma_{1}+\gamma_{2}}} > T_{1}$, where $T_{i}$ is the initial temperature of the substrate prior to the cycle. Maximum cooling power is obtained when $T_{i} = T_{H}$, and the cooling power tends to zero when $T_{i}\rightarrow T_{1}$. Here $\gamma_{1}T_{i}$ is the electronic entropy of the substrate. 
\begin{itemize}
\item{Step C: The electron fluid in the working substance is re-compressed by reducing the magnetic field quasi-statically and adiabatically, where it returns to the superconducting state at temperature $T_{4} = (T_{C}T_{*}^{2})^{1/3}$. Magnetic work $W_{2,4}$ is done.}
\item{Step D: The working substance is put in contact with the hot reservoir. Since the reservoir has a high specific heat and bandgap, the final temperature achieved can be approximated to the temperature of the hot reservoir $T_{H}$. In this process, the amount of heat transferred to the hot reservoir per unit volume is given by,
 $   Q_H = \int T dS = \frac{3 \alpha_{2}}{4} (T_4^4 - T_H^4).$
We have approximated the entropy lines for the superconducting state to be only phononic, since the electronic contribution goes to zero exponentially at low temperatures.}
\end{itemize}

Fig.~\ref{fig1} (a) illustrates this ideal process. By the first law, we have $W_{3,1}+Q_{1,2}+W_{2,4}+Q_{4,3} = 0$. Defining $W=W_{3,1}+W_{2,4}$, we have  $W=-Q_{4,3}-Q_{1,2} = Q_{H}-Q_{C}$. The Coefficient of Performance (COP) is the ratio of heat taken from the cold reservoir $Q_{C}$ to work $W$ given by,
\begin{eqnarray}
    \text{COP} =\frac{Q_{C}}{W} = \frac{t_{C}^{2}-t_{H}^{6}}{\frac{3}{2}(t_{C}^{4/3}-t_{H}^{4})-(t_{C}^{2}-t_{H}^{6})},
\end{eqnarray}
where $t_H = T_H/T_*,~t_C = T_C/T_*$. Please see Fig.~\ref{fridge}(b), where we plot the coefficient of performance as a function of $t_{H}=\frac{T_{H}}{T_{*}}$.

In the above idealized analysis, we assume an on/off type energy exchange, so heat transfer to either a hot reservoir or a cold reservoir can be made on demand, like a piston operating a heat-transfer switch. While liquid-gas refrigerators can make a good approximation to this idealized description because of their ability to be freely moved around, solid state systems do not have such freedom. Instead, we must design appropriate physics to effectively turn on and off a switch of exchanging heat with either a hot reservoir or a cold reservoir in order to make an effective solid-state refrigerator.  Below, we show that the asymmetry of heat transport between normal metals and superconductors has such a ``switch'' built in~\cite{martinez2015rectification}, which permits selectively cooling down the cold reservoir, due to the presence of an energy gap in the superconductor.  

\begin{figure}
\includegraphics[width=\linewidth]{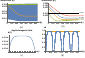}\caption{(a) Repeated cooling cycles of a junction refrigerator [see Fig.~\ref{fridge}(c)], starting from 100mK. Temperature of the working substance (blue, oscillating), the temperature variation of the substrate (orange, steady decrease with oscillations $<$1mK), and temperature of the hot reservoir (green) are shown in the figure. (b) Refrigeration action starting from different initial temperatures 0.2K, 0.1K, 0.05K and 10mK, which is the base temperature of a dilution fridge. The operating power of the refrigerator for these initial temperatures are 25 nW, 6 nW, 1.5 nW and 0.06 nW respectively, for a 1cm$\times$1cm interface.  For each case, we assume that the hot and cold reservoirs, and the working substance are in thermal equilibrium such that their temperatures are identical before the refrigerator is turned on.   (c) The applied magnetic field profile [same form for different realizations, $H(t) = (1-n)H_{c}(T_{i})+(H_{c}(0)-(1-n)H_{c}(T_{i})) \tanh(t/\tau)$ until $t=4\tau$, and then reduced symmetrically], and (d) variation of the fraction of normal metal in the working substance during repeated cooling cycles of a junction refrigerator, for $T_{i}=100$mK.
\label{fig3}}
\end{figure}

 For efficient cooling, it is desired that the working substance, when it is cold, is as much thermally isolated from the hot reservoir as possible, so that significant amount of heat is absorbed from the cold reservoir which we want to cool down. To achieve this, we consider a reservoir superconductor having a larger gap. As a result, there is significantly less back-flow of quasi-particles from the hot reservoir to the working substance when the working substance is colder (in its normal state), and there is more in-flow of heat from the cold reservoir to the working substance, since both are in their normal state. The population of quasi-particle excitations in the reservoir superconductor, which could potentially tunnel back to the working substance when it is colder, are exponentially suppressed by the presence of a large superconducting energy gap in the reservoir.   Similarly, the reverse flow of heat from the the working substance to the cold reservoir, when the working substance is hot (superconducting state) is also exponentially suppressed due to the appearance of the superconducting gap. Therefore, in each cycle, there is more heat absorbed from the cold reservoir, than the reverse flow of heat. This is further facilitated by maintaining a high magnetic field for most of the time in each cycle [see Fig.~\ref{fig3}(c)] such that the working substance spends most of its time in each cycle in the cold (normal) state. Majority of the excess heat is distributed in the phonon modes of the working substance. By increasing the volume of the working substance (and therefore its specific heat) relative to the volume of the cold reservoir, we also ensure that the temperature of the working substance increase at a relatively slow rate with the amount of heat absorbed, compared to the decrease in temperature of the cold reservoir in each cycle, adding to efficient cooling.  Electron-phonon scattering and contact with the large gap superconductor further facilitates achieving thermal equilibrium in the hot junction, by selective removal of high energy quasi-particles from the working substance. In practice, the reservoir superconductor can also come in direct contact with rest of the internal environment of a dilution refrigerator which sets the initial equilibrium temperature. Here the reservoir superconductor also provides additional thermal isolation between the working substance and the base contact, owing to the presence of a large superconducting gap in the reservoir. 

A large gap superconductor as the reservoir is also desired, if we need to produce the magnetizing B field by running a super-current in the reservoir superconductor [see Fig.~\ref{fridge}(c)], without breaking the reservoir's own superconductivity. Our choice, Niobium, as the large gap superconductor has a higher critical field ($B_{0}\sim 0.82$ T) compared to Tantalum ($B_{0}\sim 0.08$ T)~\cite{stromberg1965superconducting,eisenstein1954superconducting,milne1961superconducting}. Therefore Niobium can sustain the supercurrent that produce the magnetizing B field without breaking its own superconductivity. Niobium is also type II, and therefore it can enter a mixed state with normal vortices. It is still acceptable as the lower critical field above which Niobium enters a mixed state has been measured around 0.19 T~\cite{stromberg1965superconducting,eisenstein1954superconducting}, which is still higher than the critical field of Tantalum.  We also assume that the Kapitza coupling~\cite{pollack1969kapitza,rajauria2007electron,elo2017thermal} across the tunnel junctions can be avoided by carefully choosing the disordered tunnel barriers such that it causes phonon mismatch, and prevents phonon mediated heat transport. This is another desired feature for the experimental implementation of the refrigeration scheme presented below. Two alternate experimental implementations for our scheme that reduce phonon mediated heat transport between the junctions using suspended membranes are presented in the appendix.

\textit{Continuous adiabatic cooling} -- Here we provide a dynamical description for the gradual cooling of a substrate $N_{1}$ in contact with the working substance $S_{2}/N_{2}$, which is subsequently in contact with a hot reservoir, $S_{3}$. 
The quasiparticle tunneling across the interface and the dissipative effects determine the temperature evolution of the three regions, 1: substrate ($T_{L}$), 2: working substance ($T_{w}$), 3: hot reservoir ($T_{R}$). The adiabatic description for cooling of the working substance with dissipative effects is governed by the relation  $\frac{d\mathcal{S}_{w}}{dt} = \frac{P_{w}(t)}{T_{w}(t)}$, where
 \begin{equation}   \mathcal{S}_{w}(T_{w},t) = x_{N}(T_{w},t)\mathcal{S}_{w}^{N}(T_{w})+(1-x_{N}(T_{w},t))\mathcal{S}_{w}^{S}(T_{w}),\end{equation}
and $P_{w}$ is the net dissipative power per unit volume in the working substance, due to thermal contacts and eddy currents, and $T_{w}$ is the temperature of the working substance. Here $x_{N}(T_{w},t)$ is the fraction of normal metal in the working substance at time $t$ given by,
   \begin{equation}
         x_{N}(T_{w},t) = 1-n^{-1}\bigg(1-\frac{H(t)}{H_{c}(T_{w})}\bigg),\end{equation} 
   where $H(t)$ is the applied magnetic field and $n$ is the demagnetization factor of the material.~We set $n=5\times 10^{-4}$ for the working substance~\cite{dolcini2009adiabatic}.  Variation of $x_{N}$ for our refrigeration protocol in shown in Fig.~\ref{fig3}(d), which shows that the fraction increases from zero to one, and then falls back to zero in the proposed magnetization cycle. The dynamics of the refrigerator is described by the following set of simultaneous differential equations (assuming unit volume):
\begin{eqnarray}
C_{N_{1}}(T_{L})\dot{T_{L}} &=& -x_{N}(T_{w},t)P_{N_{1},N_{2}}^{qp}+P_{load}\nonumber\\&-&(1-x_{N}(T_{w},t))P_{N_{1},S_{2}}^{qp}\nonumber\\
C_{w}(H,T_{w})\dot{T_{w}} &=& x_{N}(T_{w},t)(P_{N_{1},N_{2}}^{qp}-P_{N_{2},S_{3}}^{qp})\nonumber\\&+&(1-x_{N}(T_{w},t))(P_{N_{1},S_{2}}^{qp}-P_{S_{2},S_{3}})\nonumber\\&+&P_{mag}+P_{eddy}.
\label{eqcont}
\end{eqnarray}
A similar dynamical equation exists for $T_{R}$, but for a large volume of the hot reservoir, and coupling to a support at fixed initial temperature, we can safely assume that $\dot{T_{R}} = 0$. The specific heat $C_{w}$ is the specific heat of the intermediate state, given by~\cite{dolcini2009adiabatic},
 \begin{equation}   C_{w}(H,T_{w}) = x_{N} C_{N}(T_{w})+(1-x_{N})C_{S}(T_{w})+C^{Lat}_{V}(H,T_{w}),\end{equation}
where \begin{eqnarray} C^{Lat}_{V}(H,T_{w}) &=& (
T_{w} H/\mu_{0}n H_{c}^{3}(T_{w}))\nonumber\\&\times&(\mathcal{S}_{w}^{N}(T_{w},0)-S^{S}_{w}(T_{w},0))^2,\end{eqnarray} corresponds to the latent heat of the phase transition. The competing cooling power is,
\begin{equation} P_{mag} = \frac{\mu_{0}}{n}T_{w}\frac{dH_{c}(T_{w})}{dT_{w}}\dot{H}.
\end{equation}  

 We treat the electron and phonon temperatures identical in Eq.~\eqref{eqcont}, since electron-phonon relaxation occurs much faster compared to adiabatic magnetization, which is a slow process.  Under this assumption, here $P_{load}$ accounts for a small heating contribution from thermal contacts by treating them as hot-spots, where lattice temperature is approximately constant in the immediate neighborhood of the contact~\cite{giazotto2006opportunities}. The heating power at each contact varies as
\begin{equation}
    P_{\text{ct}}\simeq \Sigma V_{\text{ct}}(T_{i}^{q}-T_{L}^{q}),
\end{equation} 
where to a good approximation, the dissipation is caused due to electron-phonon scattering at the contact (which sets $q=5$), with the phonon /lattice temperature in the neighborhood of the contact held fixed at the initial equilibrium temperature $T_{i}$. The volume $V_{\text{ct}}$ of the thermal hot-spot at the contact is modeled as a sphere of radius $r_{\text{ct}}$. Here $\Sigma = 2\times10^{9}~\text{WK}^{-5}\text{m}^{-3}$ is the electron phonon coupling constant for Cu. In simulations, we consider two such hot-spots with $r_{\text{ct}}\sim$ 600 nm each, accounted by $P_{load}$. This adds only a maximum heating contribution of nearly $\text{1 pW}$  at $T_{i}=200$mK, $\sim$ 0.03 pW at $T_{i}=100$mK and $\sim$ 1 fW at 50mK, which are much smaller compared to the respective cooling powers in the nW range (see Fig.~\ref{fig3}). The heating contribution further drops down with the ambient temperature $T_{i}$, set by the lowest temperature achievable in a dilution fridge. In each of the cooling curves in Fig.~\ref{fig3}(b), we assume that our refrigerator begins to function from different ambient cold temperatures $T_{i}$ achieved in a dilution fridge, and the contacts are in thermal equilibrium at this ambient temperature, $T_{i}$.

The working substance can heat up due to eddy currents introduced by the magnetic field $B=\mu_{0}x_{N}(T_{w},t)H_{c}(T_{w})$ varies as $P_{eddy}(t)= \frac{A^{2}\dot{B}^{2}}{R_{w}}$, where $A$ is the area which the normal component of the field is passing through, and $R_{w}$ is the bulk resistance of the working substance. Eddy current effects can be reduced by a factor $\propto \frac{1}{N_{w}^{2}}$, by subdividing the bulk into $N_{w}$ thin sheets. We assume $N_{w}\sim 10^{2}$ in the simulations. Note that such a laminar formation occurs naturally in the effective description of the intermediate state where the metal and superconducting phases coexist with alternating thin strips of metal  and superconducting phases, and the magnetic field lines pass only through the normal phase~\cite{rose2012introduction}. 

The quasiparticle power (energy exchange per unit time) transported between two normal metals is,
\begin{eqnarray}
    P_{N_{1},N_{2}}^{qp} &=&\frac{2}{e^{2}\mathcal{R}} \int_{0}^{\infty} E~dE(\mathcal{F}_{1}(T_{L})-\mathcal{F}_{2}(T_{w}))\nonumber\\
    &=&\frac{1}{e^{2}\mathcal{R}} \frac{\pi^{2}k_{B}^{2}}{6}(T_{L}^{2}-T_{w}^{2}).\label{pow}
\end{eqnarray}
Here $\mathcal{F}(T)$ is the Fermi-Dirac distribution at temperature $T$, $k_{B}$ is the Boltzmann constant, and $\mathcal{R}=\frac{\mathcal{R}_{s}}{\mathcal{A}}$ is the normal state resistance of the junction. The specific resistance $\mathcal{R}_{s}$ is assumed to be $2\text{M}\Omega\mu \text{m}^{2}$, and identical for both the junctions.  As expected, good energy transfer is found (going as a power law of the temperature difference) because of the density of states-matching of the two normal metals. The maximum cooling power provided by the junction can be calculated from Eq.~\eqref{pow}. As noted previously, maximum cooling power is obtained when $T_{L}=T_{i}=T_{H}$, and $T_{w}=\frac{T_{H}^{3}}{T_{*}^2}$.  Substituting, we obtain \begin{eqnarray}
    P^{\text{c}}_{\text{max}} 
    &=& \frac{\gamma_{2}}{e^{2}\mathcal{R}~\alpha_{2}} \frac{\pi^{2}k_{B}^{2}}{6}t_{H}^{2}(1-t_{H}^{4}).
\end{eqnarray}
Using the parameters mentioned in the caption of Fig.~\ref{fridge}~(c), we obtain $T_{*}=\sqrt{\frac{\gamma_{2}}{\alpha_{2}}} = 11.6$K for Tantalum. For a specific resistance $\mathcal{R}_{s}=2\text{M}\Omega\mu\text{m}^{2}$, a 10cm$\times$10cm contact has resistance $\mathcal{R}=2\times10^{-4}\Omega$, yielding the cooling power at 10mK nearly equal to 6 nW. Further maximizing $P^{\text{c}}_{\text{max}}$ over $t_{H}$ we obtain the  
optimal point of operation $t_{H}^{\text{max}} = \frac{1}{3^{1/4}}\simeq 0.76$. Since our refrigerator operates below the critical temperature $T_{c_{w}}$ of the working substance, the optimal point of operation is achievable if $t_{c_{w}}=\frac{T_{c_{w}}}{T_{*}}>t_{H}^{\text{max}}$, i.e., when
\begin{equation}
   0.746\frac{\Delta_{2}}{ k_{B}}\bigg(\frac{\alpha_{2}}{\gamma_{2}}\bigg)^{\frac{1}{2}}>1, \hspace{0.5cm}\text{using}\hspace{0.5cm}k_{B}T_{c} = \frac{\Delta}{1.764}\label{con}
\end{equation}
 from the BCS theory~\cite{bardeen1957theory}. The ideal refrigerator sketched in Fig~\ref{fridge}(c) has a $\text{COP}=1.65$ at this optimal point.
 
Similarly, the quasiparticle power exchange between a normal metal and a superconductor is given by,
\begin{eqnarray}
    &&P_{N_{1},S_{2}}^{qp} =\frac{2}{e^{2}\mathcal{R}} \int_{\Delta_{2}}^{\infty} E~dE \frac{E}{\sqrt{E^2-\Delta_{2}^2}}(\mathcal{F}_{1}(T_{L})-\mathcal{F}_{2}(T_{w}))\nonumber\\
    &&\simeq\frac{2}{e^{2}\mathcal{R}} \bigg[ \bigg(\Delta_{2}^2 K_0\left[\frac{\Delta_{2}}{k_{B}T_{L}} \right] + \Delta_{2} k_B T_L
    K_1\left[ 
    \frac{\Delta_{2}}{k_{B} T_{L}} \right]\bigg)\nonumber\\&&- \left(\Delta_{2}^2  K_0\left[\frac{\Delta_{2}}{k_{B}T_{w}}\right] + \Delta_2 k_B T_w K_1\left[
     \frac{\Delta_{2}}{k_{B} T_{w}} \right]\right)\bigg].
\end{eqnarray}
Here $\Delta_{2}$ is the energy gap of $S_{2}$, and $K_{0,1}(x)$ are modified Bessel functions of order 0 and 1. A similar relation can be found for $P_{N_{2},S_{3}}^{qp}$. In pursuing the integrals, we have assumed low temperatures such that the integrals are effectively approximated using Laplace transformations. For large $\Delta_{2}/k_B T_{L,w}$, the asymptotic expansion of Bessel function, $K_n(x) \sim e^{-x} \sqrt{\pi/2x}$, insures an exponential cut-off of the transport between the N/S junction, acting as the desired switch.

The heat exchange between the two superconducting elements has two contributions, the quasiparticle power exchange, $P_{S_2,S_3}^{qp}$ and a term depending on the Josephson phase, $\phi=\phi_{R}-\phi_{w}$,~ $P_{S_2,S_3}^{\phi}$~\cite{maki1965entropy,giazotto2012josephson}. Here $\phi_{R},~\phi_{w}$ are respectively the phase of the supercondcuting BCS wavefunctions of $R$ and $w$. The quasiparticle tunneling power across the $S_2/S_3$ junction is approximated~\cite{golubev2013heat},
\begin{eqnarray}
    &&P_{S_{2},S_{3}}^{qp}
    \simeq\frac{1}{e^{2}\mathcal{R}} \frac{\sqrt{2 \pi} \Delta_{3}^{5/2}}
  {\sqrt{\Delta_{3}^{2} - \Delta_{2}^{2}}} \bigg(\sqrt{k_{B} T_{w}} e^{-\Delta_{3}/(k_B T_{w})} \nonumber\\&&\times\cosh\bigg(\frac{\hbar\dot{\phi}}{2k_{B}T_{w}}\bigg)-
   \sqrt{k_{B} T_{R}} e^{-\Delta_{3}/(k_{B} T_{R})}\bigg),
\end{eqnarray}
where we have assumed that the difference $\Delta_{3} - \Delta_{2}$ is much bigger than the thermal energies of quasiparticles which help reduce the back-flow of heat from the reservoir to the working substance. The magnitude of the $\phi$ dependent term is always smaller than  $P_{S_{2},S_{3}}^{qp}$ and is given by $P_{S_{2},S_{3}}^{\phi}=-\Delta_{2}/\Delta_{3}P_{S_{2},S_{3}}^{qp}\cos{\phi}$~\cite{golubev2013heat}. In the examples considered, we have taken $T_{c_{w}} = 4.48$K for Tantalum, $T_{c_{R}} = 9.29$K for Niobium. The refrigerator operates below $0.1~T_{c_{w}}$ in the examples presented, where the superconducting gap remains constant at its zero temperature value. In Fig.~\ref{fig3}, we have set $\phi = 0$, and $\dot{\phi}=0$. In general, the relative phase between the superconductors provides another control knob in the problem, and is significant in determining the cooling power when the magnetizing cycles are applied faster than the thermal relaxation time of the Josephson junction~\cite{fornieri2017towards,martinez2014coherent}.  

\textit{Discussions} -- We proposed a cyclic superconducting refrigerator using the principle of adiabatic magnetization of a superconductor. The refrigerator action is similar to a conventional kitchen refrigerator. Here, the working fluid is the electron gas in a superconductor switching between normal (expanded) and superconducting (compressed) states in an applied magnetic field. Substantial cooling down of a substrate is predicted, as depicted in Fig.~\ref{fig3}(b) for different equilibrium initial temperatures of the refrigerator. Although we discussed a refrigeration scheme in which conventional s-wave superconductivity and the BCS description holds, we note that similar adiabatic cooling effects can be achieved in high-temperature superconductors as well. For example, cooling by adiabatically increasing the super-current in a high temperature superconductor has been studied by Svidzinsky in Ref.~\cite{svidzinsky2002possible}, which could be an alternate way of achieving single shot adiabatic cooling with superconductors~\cite{dolcini2009adiabatic}. It should be noted that a high temperature superconductor may increase the operating temperature of the refrigerator, but cooling down a normal metal may still be more efficient in the low temperature regime, as the phonon entropy ($\propto T^{3}$), and electron phonon scattering ($\propto T^{5}$) starts to dominate at high temperatures, effectively nullifying any cooling effect in the normal metal from electron-mediated transport phenomenon at the interface.

We conclude by noting that many variations of our proposal are possible.  For example, if another set of metal/superconductor junctions is placed on the other side of the metal to be cooled, then out-of-phase double-action refrigeration is possible, where one side continues to cool the metal down while the other side is heating up and ejecting its excess heat. Our solid state refrigeration technique can be very effective for achieving significant cooling in superconducting circuits, and for applications such as superconducting single photon detectors~\cite{gol2001picosecond} and sensors~\cite{giazotto2008ultrasensitive}. 

\textit{Acknowledgements} -- Work by SKM and ANJ was supported by the U.S. Department of Energy (DOE), Office of Science, Basic Energy Sciences (BES), under Award No. DE-SC0017890. Work by FG was supported by the European Research Council under the European Union’s Seventh Framework Programme (FP7/20072013)/ERC Grant Agreement No. 615187-COMANCHE. This research was supported in part by the National Science Foundation under Grant No. NSF PHY-1748958 (KITP program QTHERMO18). We thank Jukka Pekola, Björn Sothmann, Rafael Sánchez, Matthew LaHaye, and Britton Plourde for discussions, and also Elia Strambini for suggesting the possible method of injecting a super-current in the reservoir superconductor to realize the magnetizing B field. 
\appendix 
\section{Alternative implementation schemes for the refrigerator}
  \begin{figure*}
\includegraphics[width=\linewidth]{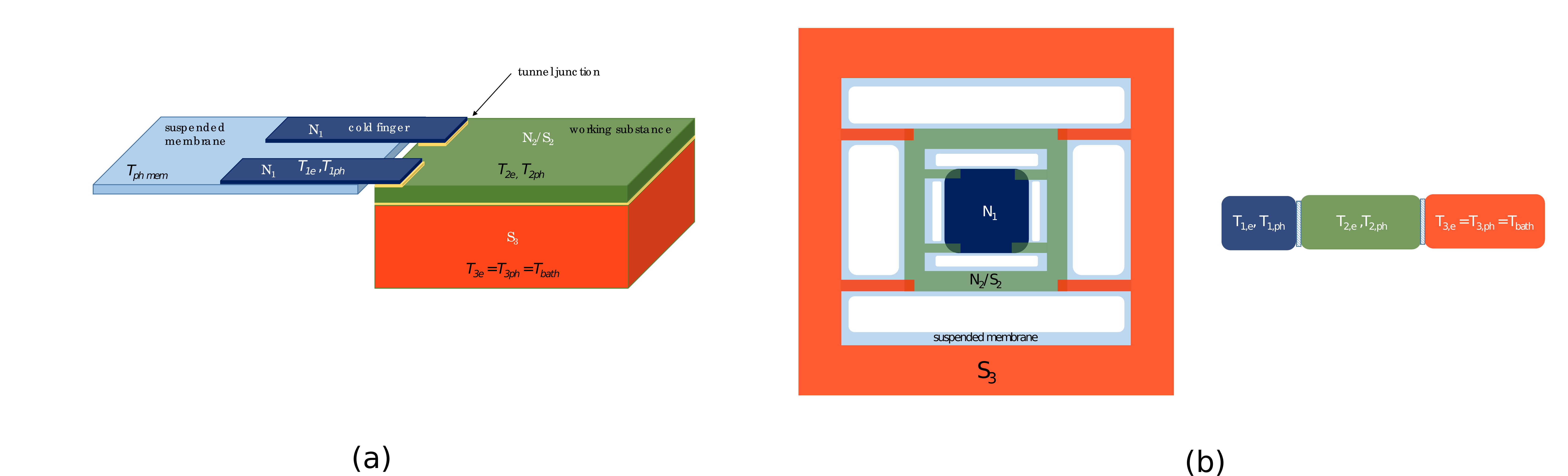}\caption{(a) We assume that the phonon mediated heat transport (Kapitza coupling) between the working substance and the hot reservoir can be suppressed by choosing the tunnel barrier appropriately, while the substrate is isolated by suspending it as a membrane to reduce Kapitza coupling. (b) In this architecture, the Kapitza coupling across both the junctions is reduced by suspending the working substance and the substrate in separate membranes. 
\label{figSM}}
\end{figure*}
 Here we propose two alternative implementation schemes for the refrigeration protocol discussed in the letter. Both use a suspended membrane to reduce the Kapitza coupling
 in order to inhibit phonon thermal transport between interfaces (See Fig.~\ref{figSM}). 
 
 The Kapitza coupling for an interface between materials $j$ and $k$, with phonon temperatures $T_{ph_{j}}$ and $T_{ph_{k}}$ is given by~\cite{pollack1969kapitza,rajauria2007electron,elo2017thermal},
 \begin{equation}
     P_{ph_{j},ph_{k}} = KA(T_{ph_{j}}^{4}-T_{ph_{k}}^{4}),
 \end{equation}
 where $A$ is the area of the interface, and $K$ is the coupling$~\sim 200~$W m$^{-2}$ K$^{-4}$, for typical  metal interfaces.  We stress that in general, different temperatures for electrons and phonons can be investigated in this scheme, as marked in the figures. Negligible Kapitza coupling, and fast electron-phonon interaction relative to the adiabatic magnetization process ensures that we can essentially treat the electron/phonon temperatures the same during the quasi-static operation of the refrigerator, as described in Eq.~\eqref{eqcont} of the main text.
\bibliography{ref}
   \end{document}